\documentclass[prl,twocolumn,amsmath,amssymb]{revtex4-2}

\usepackage{graphicx}
\usepackage{dcolumn}
\usepackage{bm}
\usepackage{mathrsfs}
\usepackage{subfigure}
\usepackage{natbib}
\usepackage{amsmath}
\usepackage{amssymb}
\usepackage{Jonasmacros}
\usepackage{mathptmx}
\usepackage{pifont}
\usepackage{times}
\usepackage{txfonts}

\usepackage[usenames,dvipsnames,svgnames]{xcolor}
\usepackage[colorlinks,citecolor=blue,linkcolor=Fuchsia]{hyperref}

\date{\today} 
\begin{document}

\title{Charge and Spin Dynamics and Enantioselectivity in Chiral Molecules}

\author{J. Fransson}
\email{Jonas.Fransson@physics.uu.se}
\affiliation{Department of Physics and Astronomy, Box 516, 751 21, Uppsala University, Uppsala, SWEDEN}




\begin{abstract}
Charge and spin dynamics is addressed in chiral molecules immediately after the instantaneous coupling to an external metallic reservoir. It is shown how a spin-polarization is induced in the chiral structure as a response to the charge dynamics. The dynamics indicate that chiral induced spin selectivity is an excited states phenomenon which, in the transient regime partly can be captured using a simplistic single particle description, however, in the stationary limit definitively shows that electron correlations, e.g., electron-vibration interactions, crucially contribute to sustain an intrinsic spin anisotropy that can lead to a non-vanishing spin selectivity. The dynamics, moreover, provide insight to enantiomer separation, due to different acquired spin-polarizations.
\end{abstract}
\maketitle

During the course of more than the past two decades, we have learnt that electron spin selective processes are intimately associated with chirality \cite{Science.283.814,Science.331.894}. The effect can partly be understood as emerging from a combination of structural chirality, spin-orbit interactions, and non-equilibrium conditions \cite{PhysRevE.98.052221,PhysRevB.99.024418,NJP.20.043055,JPhysChemC.123.17043,PhysRevB.85.081404(R),PhysRevLett.108.218102,PNAS.111.11658,JPhysChemC.117.13730,PhysRevB.93.075407,PhysRevB.93.155436,ChemPhys.477.61,NanoLett.19.5253,JPhysChemLett.9.5453,JPhysChemLett.9.5753,JChemTheoryComput.16.2914,JPhysChemLett.10.7126,CommunPhys.3.178,NewJPhys.22.113023,PhysRevB.102.035431,PhysRevB.102.214303,PhysRevB.102.235416,JChemPhys.154.110901,JACS.143.14235,NanoLett.21.6696,NanoLett.21.8190,PhysRevB.104.024430}.
Despite that non-equilibrium conditions, which define the measurements, for instance, light exposure \cite{Science.283.814,Science.331.894,PNAS.110.14872,NanoLett.14.6042,NatComms.7.10744,AdvMat.30.1707390,JPhysChemLett.9.2025,Chirality.33.93}, local probing techniques \cite{NanoLett.11.4652,JPhysChemLett.11.1550,AdvMater.28.1957,ACSNano.14.16624}, transport \cite{NatComms.4.2256,JPhysChemLett.10.1139,JPhysChemLett.11.1550} and different types of Hall measurements \cite{NatComms.7.10744,NatComms.8.14567,AdvMat.30.1707390,PhysRevLett.124.166602,PhysRevLett.127.126602}, are present, many theoretical accounts of the effect are based on the transmission properties of chiral molecules embedded in a given environment \cite{PhysRevE.98.052221,PhysRevB.99.024418,NJP.20.043055,JPhysChemC.123.17043,PhysRevB.85.081404(R),PhysRevLett.108.218102,JPhysChemC.117.13730,PhysRevB.93.075407,PhysRevB.93.155436,ChemPhys.477.61,JPhysChemLett.9.5453,JPhysChemLett.9.5753,JChemTheoryComput.16.2914,CommunPhys.3.178,NewJPhys.22.113023}. While the transmission pertains to the linear response regime, hence, the ground state properties of the molecule, it is also often typically the result of a single particle description which under stationary condition cannot account for the excited states properties that underlie spin selectivity in chiral molecules.

Recent theoretical developments very clearly point towards the necessity to consider chiral induced spin selectivity from an excited states point of view \cite{JPhysChemLett.10.7126,NanoLett.20.7077,PhysRevB.102.214303,PhysRevB.102.035431,PhysRevB.102.235416,NanoLett.21.3026,JACS.143.14235,NanoLett.21.6696,NanoLett.21.8190,PhysRevB.104.024430}, stressing the vital role of electronic correlations.
In this context, excited states refer to, e.g., virtual excitations, temporal fluctuations induced electron dispersion within the spectrum, and thermally induced vibrational excitations of the molecule which couple to the electron and, hence, strongly modify the electronic structure. There might, however, be other sources for exciting the molecule.
It was shown that, e.g., Coulomb \cite{JPhysChemLett.10.7126,NanoLett.20.7077,JACS.143.14235} and electron-vibration \cite{PhysRevB.102.035431,PhysRevB.102.235416,JChemPhys.154.110901,NanoLett.21.3026} interactions, as well as polarons \cite{PhysRevB.102.214303}, photoexcitations \cite{NanoLett.21.6696}, time-dependence \cite{NanoLett.21.8190}, and dissipation \cite{PhysRevB.104.024430} generate the exchange necessary to create measurable effects regarding chiral induced spin selectivity.

In this context, an obvious question is how a spin-polarization can be generated and stabilized in a molecular structure which is in a closed shell configuration when isolated from the surrounding environment. Recent experiments demonstrate that a measurable spin-polarization can be obtained whenever chiral molecules are interfaced with metallic surfaces \cite{NatComms.8.14567,Molecules.25.6036,PNAS.114.2474,NanoLett.19.5167}. Through the anomalous Hall effect, chiral molecules were, for instance, shown to control the magnetism in thin Co layers \cite{NatComms.8.14567,Molecules.25.6036} and enantiomer separation on non-magnetic metals \cite{PNAS.114.2474}, whereas Yu-Shiba-Rusinov states \cite{ProgTheorPhys.40.435,ActaPhysSin.21.75,JETPLett.2.85} were observed in the vicinity of chiral molecules on the surface of superconducting NbSe$_2$ \cite{NanoLett.19.5167}.
Related to these observations are also the results showing strongly enantiomer dependent binding energies on ferromagnetic metals \cite{Science.360.1331,ApplPhysLett.115.133701,JChemPhysB.123.9443,AdvMater.31.1904206,JPhysChemC.125.17530,JPhysChemLett.12.5469}. Theoretically, enantiomer separation was addressed in Refs. \citenum{NanoLett.20.7077,NanoLett.21.3026} for molecules in contact with ferromagnetic metal, based on descriptions in which the effective electronic exchange plays a crucial role in the magnetic response. On the other hand, while excited states appear to be crucial, the question of how spin-polarization emerges in chiral molecules when in contacted with a metal remains open.

In this Letter, it is shown that a finite spin-polarization is dynamically generated in chiral molecules as a response to the charge dynamics when interfaced with a metal. The dynamics indicate that chiral induced spin selectivity is an excited states phenomenon which, in the transient regime partly can be captured using a simplistic single particle description, while such a description is not sufficient in the stationary limit. The latter statement is founded on that the spin-polarization eventually vanishes which, in turn, implies that any mechanism that can sustain an immanent spin anisotropy has to account for excited states properties, e.g., electron-electron or electron-vibration interactions. In the subsequent discussion it is, therefore, shown that, by adding electron-vibration interactions, the transient spin fluctuations are stabilized and developed into a finite spin-polarization as the stationary limit is approached. Finally, when interfacing the different enantiomers on a ferromagnet, the results presented herein provide fundamental clues for a comprehensive picture of enantiomer separation.


The presented discussion is based on simulations of idealized chiral models of realistic, e.g., $\alpha$-helix, oligopeptides, polyalanines, and helicene. In this context, it is important that while the specific details of the molecules used in the experiments strongly vary, the salient properties such as chirality, spin-orbit interaction, and interface to a metal are captured within this model. The model was previously proposed in order to point out the importance of electronic correlations originating from Coulomb \cite{JPhysChemLett.10.7126} and electron-vibration \cite{PhysRevB.102.235416,NanoLett.21.3026} interactions. Here, such mechanisms are first excluded for sake of highlighting the dynamics of excited states as a fundamental component in the charge redistribution and accompanied spin-polarization immediately after interfacing the molecule with the metal. Then, in the subsequent discussion it is shown that electron-vibration interactions stabilize a non-vanishing spin-polarizations also in the stationary limit, hence, emphasizing that the phenomenon of chiral induced spin selectivity is an excited states effect.


\begin{figure}[t]
\begin{center}
\includegraphics[width=0.75\columnwidth]{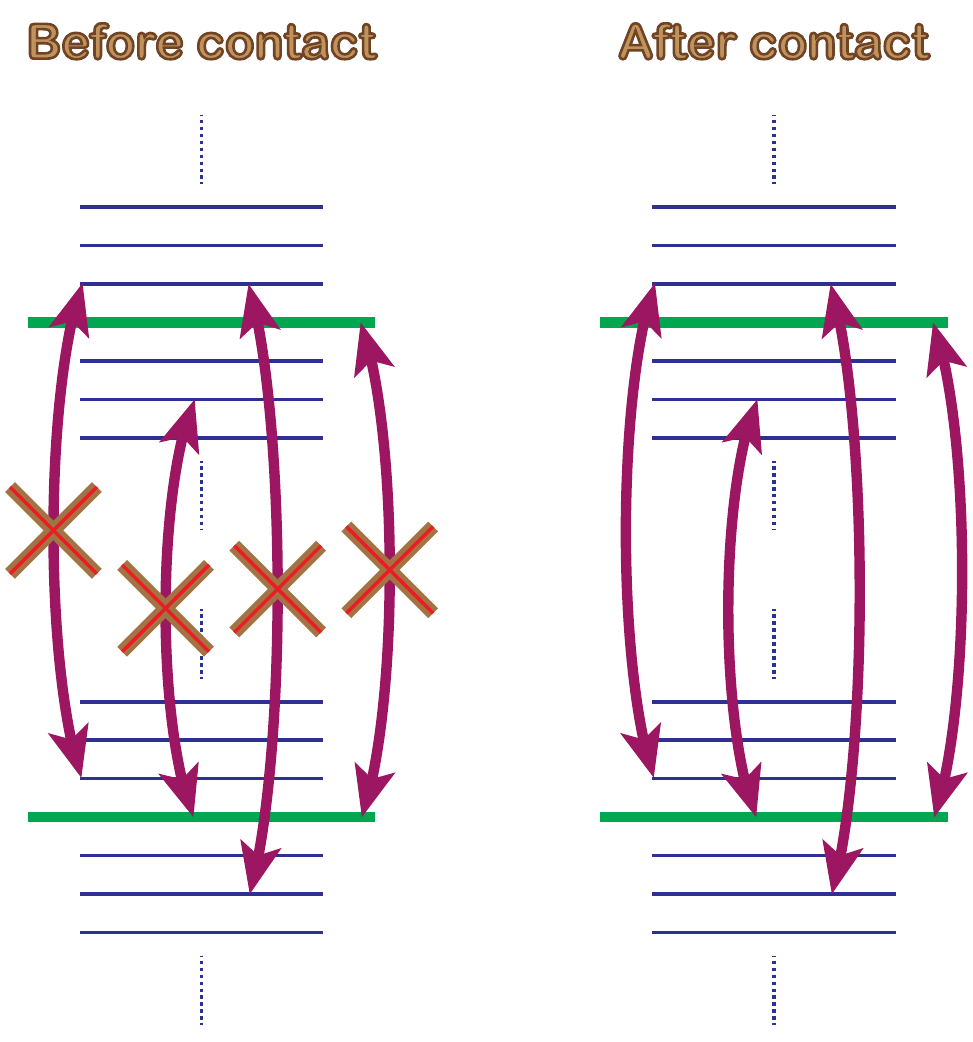}
\end{center}
\caption{Schematic drawing of the electronic processes in the chiral molecule before and after contact with the metal. Bold (faint and dotted) lines represent the main (vibrationally induced) electronic levels, arrows signify a few possible transitions between the states which, before contact are not allowed by orthogonality whereas these are allowed after making contact with the metal since the presence of the metal breaks up the orthogonality between these states.}
\label{fig-Schematic}
\end{figure}

The drawings in Fig. \ref{fig-Schematic} illustrates the electronic processes within the molecule before and after making contact with the metal. Transitions (arrows) between the main electronic states (bold), the vibrationally induced states (faint and dotted), or the main and vibrationally induced states are not allowed (crosses) by orthogonality of the electronic spectrum. By contrast, the presence of the metal breaks up the orthogonality of the spectrum which allows all transitions that were forbidden before contact.

The simulations are based on a model of a chiral structure comprising a set of $\mathbb{M}=M\times N$ nuclear coordinates $\bfr_m=(a\cos\varphi_m,a\sin\varphi_m,c_m)$, $\varphi_m=2\pi(m-1)/(\mathbb{M}-1)$, and $c_m=c\varphi_m/2\pi$, where $a$ and $c$ define the radius and length, respectively, of the helix of $M$ laps and $N$ nuclei per lap. The molecule is described by the single-electron levels $\bfepsilon=\diag{\dote{1},\dote{2},\ldots,\dote{\mathbb{M}}}{}$, representing the energy levels at $\bfr_m$, associated with the electron creation and annihilation spinors $\psi^\dagger_m$ and $\psi_m$, respectively. For equivalent sites, it is justified to assume that $\dote{m}=\dote{0}$ for all $m$. Nearest neighbors interact via elastic and inelastic hopping with rates $t_0$ and $t_1$, respectively, while next-nearest neighbors interact via the effective static and non-static spin-orbit interaction with rates $\lambda_0$ and $\lambda_1$, through processes of the type $i\psi^\dagger_m\bfv_m^{(s)}\cdot\bfsigma\psi_{m+2s}$, $s=\pm1$, where $\bfv_m^{(s)}=\hat\bfd_{m+s}\times\hat\bfd_{m+2s}$, defines the chirality in terms of the unit vectors $\hat\bfd_{m+s}=(\bfr_m-\bfr_{m+s})/|\bfr_m-\bfr_{m+s}|$, such that different enantiomers are represented by the sign ($\pm$) of the chirality. It is justified to couple the chirality with the spin-orbit interaction mechanism, since the geometry of the structure inevitably can be related to an intrinsic electric field. The notation $\sigma^0$ and $\bfsigma$ refer to the identity and vector of Pauli matrices, respectively. The nuclear, or, molecular vibrations are captured in the coherent vibrational mode $\omega_0$, which is created and annihilated by the phonon operators $b^\dagger$ and $b$, respectively. A Hamiltonian model can be written $\Hamil_\text{mol}=\Hamil_0+\Hamil_1=\Psi^\dagger[\bfH_0+\bfH_1(b+b^\dagger)]\Psi$, where $\Psi=(\psi_1,\psi_2,\ldots,\psi_\mathbb{M})^t$ and
\begin{subequations}
\label{eq-model}
\begin{align}
\label{eq-H0}
\Hamil_0=&
	\dote{0}
	\sum_{m=1}^\mathbb{M}
		\psi^\dagger_m\psi_m
	+\omega_0b^\dagger b
	-
	t_0
	\sum_{m=1}^{\mathbb{M}-1}
		\Bigl(
			\psi^\dagger_m\psi_{m+1}+H.c.
		\Bigr)
\nonumber\\&
	+
	\lambda_0
	\sum_{m=1}^{\mathbb{M}-2}
		\Bigl(
			i\psi^\dagger_m\bfv_m^{(+)}\cdot\bfsigma\psi_{m+2}+H.c.
		\Bigr)
	,
\\
\label{eq-H1}
\Hamil_1=&
	-
	t_1
	\sum_{m=1}^{\mathbb{M}-1}
		\Bigl(
			\psi^\dagger_m\psi_{m+1}+H.c.
		\Bigr)
		(b+b^\dagger)
\nonumber\\&
	+
	\lambda_1
	\sum_{m=1}^{\mathbb{M}-2}
		\Bigl(
			i\psi^\dagger_m\bfv_m^{(+)}\cdot\bfsigma\psi_{m+2}+H.c.
		\Bigr)
		(b+b^\dagger)
	.
\end{align}
\end{subequations}

The properties of the metal to which the molecule is connected are captured by the parameter $\bfGamma=\Gamma(\sigma^0+p\sigma^z)/2$, representing the coupling between the nuclear site 1 and the metal. Here, $\Gamma=2\pi\sum_{\bfk\sigma}|v_{\bfk\sigma}|^2\rho_\sigma(\dote{\bfk})$ accounts for the spin-dependent hybridization $v_{\bfk\sigma}$ and spin-density $\rho_\sigma(\dote{\bfk})$ of the electrons in the metal, whereas $|p|\leq1$ denotes the effective spin-polarization of the coupling.


The time-evolution of the electronic structure of the molecule can be related to the time-dependent Green function $\bfG_{mn}(t,t')=\eqgr{\psi_m(t)}{\psi^\dagger_n(t')}$ through, e.g., the charge $\av{n_m(t)}=(-i){\rm sp}\bfG^<_{mm}(t,t)$ and spin-moment $\av{\bfs(t)}=(-i){\rm sp}\bfsigma\bfG^<_{mm}(t,t)/2$, where $\bfG^{</>}(t,t')$ is proportional to the density of occupied/unoccupied electron states. Here, ${\rm sp}$ denotes the trace over spin 1/2 space.

\begin{figure*}[t]
\begin{center}
\includegraphics[width=\textwidth]{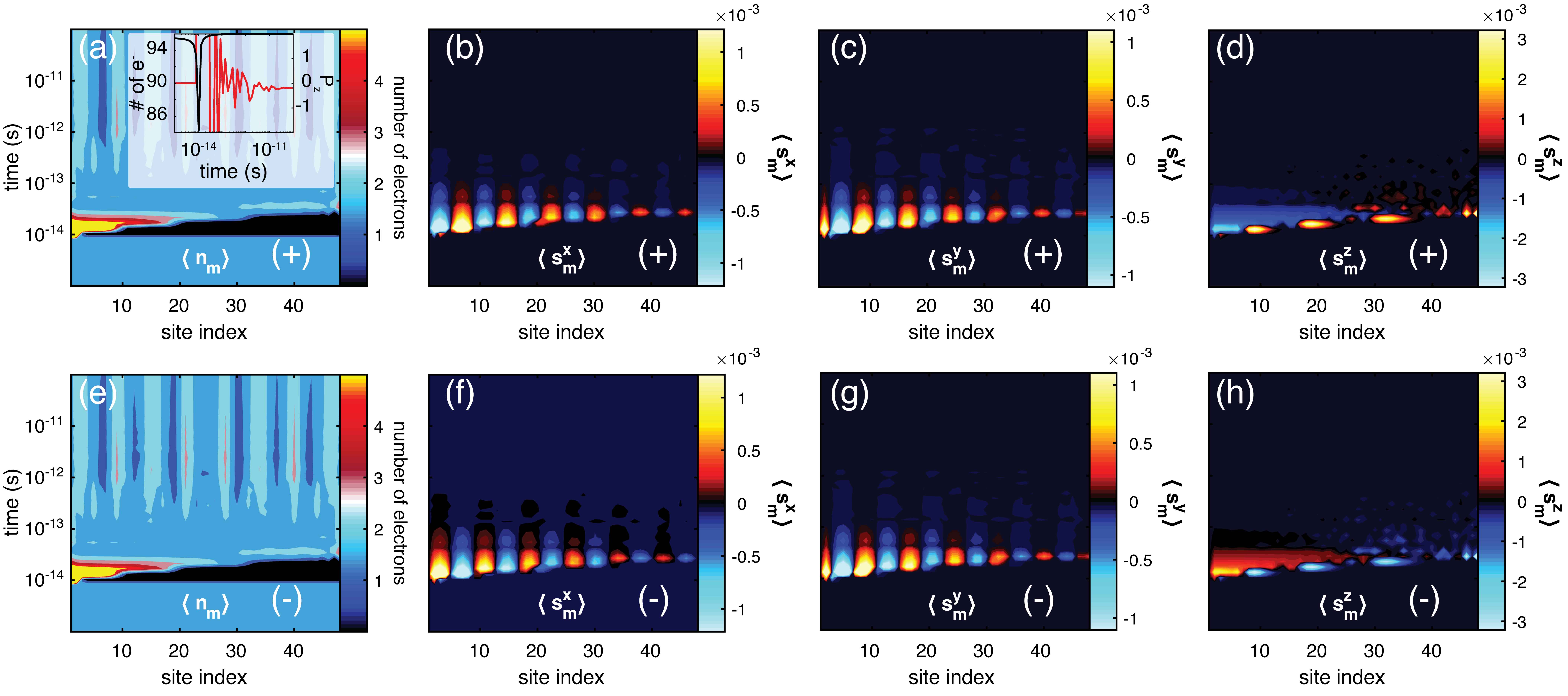}
\end{center}
\caption{Single electron picture of the charge and spin dynamics in a chiral molecule ($6\times8$) before and after making instantaneous contact with a metal at time $T_0=10$ fs. The spatially resolved charge $\av{n_m}$ (a), (e), and spin projections $\av{s_m^x}$ (b), (f), $\av{s_m^y}$ (c), (g), $\av{s_m^z}$ (d), (h), are simulated for positive ($+$) (a) -- (d) and negative ($-$) (e) -- (h) helicity. Inset in panel (a) shows the total number of electrons (black) and charge polarization (red) in the molecule as function of time. Parameters used are $a=5$ \AA, $c=235$ \AA, $\dote{0}-\mu=-4$, $\lambda_0=1/1,000$, and $\Gamma=1/100$, in units of $t_0=1$ eV, simulated at $T=300$ K.}
\label{fig-NS}
\end{figure*}

As the main interest here lies on the time-dependence of the molecular properties immediately after the time $T_0$ when the molecule is interfaced with the metal, the interaction with the metal is treated with a time-dependent hybridization $v_{\bfk\sigma}(t)=v_{\bfk\sigma}\theta(t-T_0)$, where $\theta(t)$ is the Heaviside step function. In the view of this time-dependent interaction, the equation of motion for the static ($\bfH_1=0$) Green function can be written as the Dyson equation
\begin{subequations}
\begin{align}
\label{eq-Dyson}
\bfG(t,t')=&
	\bfg(t,t')
	+
	\int
		\bfg(t,\tau)\bfSigma(\tau,\tau')\bfG(\tau',t')
	d\tau
	d\tau'
	,
\\
\label{eq-lead}
\bfSigma(t,t')=&
	\sum_\bfk\bfv_\bfk(t)\bfg_\bfk(t,\tau)\bfv^\dagger_\bfk(t')
	.
\end{align}
\end{subequations}
The bare Green functions $\bfg(t,t')=(-i){\rm T}e^{-i\bfH_0(t-t')}$ and $\bfg_\bfk(t,t')=(-i){\rm T}\sigma^0e^{-i\dote{\bfk}(t,t')}$ carry trivial time-dependencies, and alluding to the wide band limit for the electronic band in the metal \cite{PhysRevB.50.5528}, the lesser self-energy can be simplified into
\begin{align}
\bfSigma^<(t,t')=&
	i
	\bfGamma\theta(t-T_0)\theta(t'-T_0)
	\int f(\omega)e^{-i\omega(t-t')}\frac{d\omega}{2\pi}
	,
\end{align}
where $f(\omega)=f(\omega-\mu)$ is the Fermi-Dirac distribution function defined at the chemical potential $\mu$ of the metal. Thanks to this simplification, and the relation $\bfG^<=\bfG^r\bfSigma^<\bfG^a$, where $\bfG^{r/a}$ is the retarded/advanced Green function, the dressed lesser and retarded/advanced Green functions for the molecule becomes
\begin{subequations}
\begin{align}
\bfG^<(t,t')=&
	i
	\int_{T_0}^\infty
		f(\omega)
		\bfG^r(t,\tau)
		\bfGamma
		\bfG^a(\tau',t')
		e^{-i\omega(\tau-\tau')}
	d\tau
	d\tau'
	,
\label{eq-Gless}
\\
\bfG^{r/a}(t,t')=&
	(\mp i)\theta(\pm t\mp t')
	e^{-i\bfH_0(t-t')\mp\bfGamma(t,t')/2}
	,
\label{eq-Gra}
\end{align}
\end{subequations}
where the expression in Eq. \eqref{eq-Gless} is valid for $t,t'>T_0$, whereas $\bfGamma(t,t')=\bfGamma\int_{t'}^t\theta(s-T_0)ds$.

Effects from inelastic scattering are included by repeating Eq. \eqref{eq-Dyson}, defining a dressed Green function $\mathbb{G}$ and self-energy $\bfSigma_\text{vib}(t,t')=i\bfH_1\bfg(t,t')d(t,t')\bfH_1$, where $d(t,t')=2{\rm T}\sin\omega_0(t-t')$ is the propagator for the nuclear vibrations. However, since the aim is to provide a simple description of the charge and spin dynamics and a mechanism which eventually leads to a non-vanishing spin-polarized stationary state, as discussed in Ref. \citenum{NanoLett.21.3026}, the full Dyson equation for $\mathbb{G}$ is reduced to the Markovian approximation, in which the self-energy becomes time-independent. In this form, the resulting retarded Green function can be written on the form
\begin{align}
\mathbb{G}^r(t,t')=&
	(-i)\theta(t-t')
	e^{-i(\bfH_0+\bfSigma_\text{vib})(t-t')\mp\bfGamma(t,t')/2}
	,
\label{eq-Gr}
\end{align}
in which expression $\bfSigma_\text{vib}=\bfH_1\bfL\bfH_1$, where $\bfL$ is a diagonal matrix where the entries are defined by the electron-phonon loop
\begin{align}
L_{mm}=&
	\frac{n_B(\omega_0)+1-f(\dote{m})}{\dote{m}+\omega_0-i/\tau_\text{ph}}
	+
	\frac{n_B(\omega_0)+f(\dote{m})}{\dote{m}-\omega_0-i/\tau_\text{ph}}
	,
\end{align}
and $n_B(\omega)$ is the Bose-Einstein distribution function, whereas $\tau_\text{ph}$ defines an intrinsic vibrational life-time. This life-time arises from vibration-vibration and vibration-electron interactions and reflects the conditions of the environment.


\begin{figure*}[t]
\begin{center}
\includegraphics[width=\textwidth]{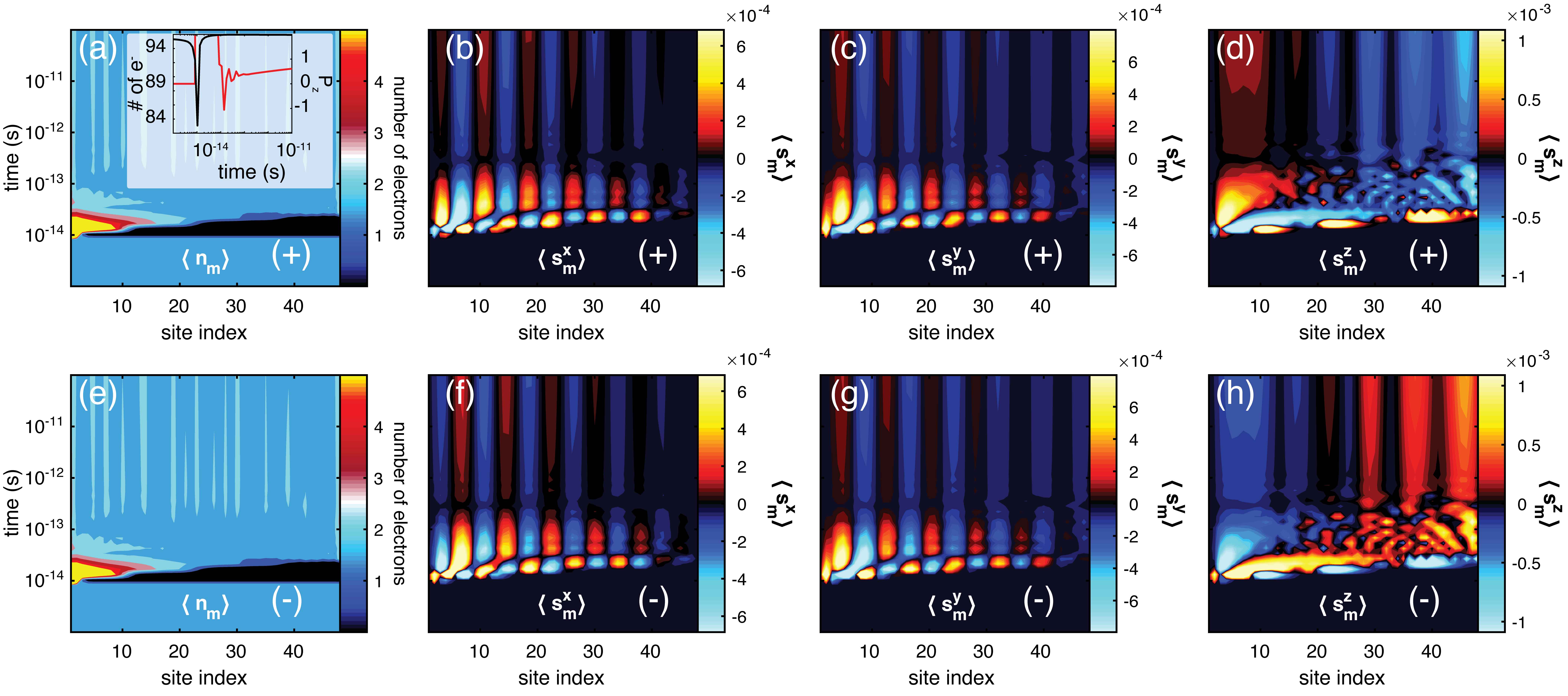}
\end{center}
\caption{Charge and spin dynamics in a vibrating chiral molecule ($6\times8$) before and after making instantaneous contact with a metal at time $T_0=10$ fs. The spatially resolved charge $\av{n_m}$ (a), (e), and spin projections $\av{s_m^x}$ (b), (f), $\av{s_m^y}$ (c), (g), $\av{s_m^z}$ (d), (h), are simulated for positive ($+$) (a) -- (d) and negative ($-$) (e) -- (h) helicity. Inset in panel (a) shows the total number of electrons (black) and charge polarization (red) in the molecule as function of time. Parameters used are $t_1=1/25$, $\lambda_1=1/10,000$, $\omega_0=1/10,000$, and $1/\tau_\text{ph}=1/100$, in units of $t_0=1$ eV, while other parameters are as in Fig. \ref{fig-NS}.}
\label{fig-vibNS}
\end{figure*}

In a previous discussion, it was shown that the molecular structure given in $\bfH_0$, Eq. \eqref{eq-H0}, carries necessary components that lift the spin-degeneracy, arising from a confluence of the spin-orbit interactions and the chiral geometry \cite{JPhysChemLett.10.7126}. However, the structure has to include at least four nuclear sites and be under non-equilibrium conditions, of which the latter was previously considered to be provided by an external voltage bias. The lack of a coupling between the internal and dissipative degrees of freedom does, nonetheless, not allow the structure to maintain a stationary spin-polarization when held in equilibrium with one end attached to a metal, see Ref. \citenum{NanoLett.21.3026}. Such a mechanism is provided by the coupling between electronic and vibrational degrees of freedom in $\bfH_1$, Eq. \eqref{eq-H1}, where the spin-independent and -dependent components broaden and introduce a spin exchange, respectively, in the spectrum. Hence, it is shown that the elastic spectrum provided through $\bfH_0$ only captures strictly non-equilibrium properties, while the stationary properties emerges from $\bfH_1$. The discussion stresses that while electron correlations are necessary for an adequate description, the single electron picture provides crucial insight into the emergence of the spin symmetry breaking.


\begin{table}[b]
\caption{Model parameters used for the simulations in units of $t_0=1$ eV.}
\label{tab-parameters}
\begin{tabular}{l|cccccc}
 & $t_1$ & $\lambda_0$ & $\lambda_1$ & $\dote{0}-\mu$ & $\omega_0$ & $\Gamma$
\\\hline
Static molecule & 0 & $10^{-3}$ & 0 & -4 & -- & 0.01 \\ 
Vibrating molecule & 0.04 & $10^{-3}$ & $10^{-4}$ & -4 & $10^{-4}$ & 0.01 \\
\end{tabular}
\end{table}

Within the given model, Eq. \eqref{eq-model}, the magnetic moment of the molecule vanishes whenever the molecule is isolated from the metallic environment. At the time when the molecule is attached to the metal, the molecule is set into a strongly non-equilibrium state where electrons are passing through the interface between the molecule and metal, rendering a redistribution of charge in the molecule. These processes are captured very well already at the single electron level ($\bfH_1=0$), see Fig. \ref{fig-NS} (a), (e), where a spatially resolved time-evolution of the molecular charge, $\av{n_m}$, is shown for a ($6\times8$ sites, $\dote{m}=\dote{0}$, for all $m$) chiral molecule with (a) positive and (e) negative chirality immediately after the instantaneous attachment of the molecule to a normal metal, with the parameters summarized in Table \ref{tab-parameters}. The plots show that the time-evolution of the charge distributions of two enantiomers are the same. The plots also show the strong fluctuations of the charge due to the abrupt changes in the environment. For instance, immediately after contact with the metal, the total molecular charge is reduced and restored shortly thereafter accompanied with a strongly fluctuating charge polarization which eventually vanes, see inset of Fig. \ref{fig-NS} (a), as a result of electrons flowing between the molecule and metal.

The charge motion in the chiral molecule generates an accompanied spin-polarization which evolves in both time and space, see Fig. \ref{fig-NS} (b)--(d), (f)--(h), showing the projections of $\av{\bfs_m}$ for the two enantiomers. The plots illustrate that there is a finite time frame during which the spin-polarization is significant, after which it vanes and eventually vanishes (between 0.1 ps and 1 ps after contact).

Concerning the spin-polarization, there are two distinctive features which are clearly illustrated in Fig. \ref{fig-NS}. First, the transverse projections, $\av{s^{x,y}_m}$, indicate anti-ferromagnetic spin configurations resulting in no, or little, net moment perpendicular to the length direction of the molecule. This is in sharp contrast to the longitudinal projection $\av{s^z_m}$, indicating a more uniform distribution in the structure. Second, the two enantiomers acquire opposite spin-polarizations such that the spin moments $\av{\bfs_m}_\pm=(\av{s^x_m}_\pm,\av{s^y_m}_\pm,\av{s^z_m}_\pm)$ in the $\pm$ enantiomers, respectively, are related by rotating the spins around the $y$-projection, that is, $(\av{s^x_m}_-,\av{s^y_m}_-,\av{s^z_m}_-)=(-\av{s^x_m}_+,\av{s^y_m}_+,-\av{s^z_m}_+)$.

Inclusion of molecular vibrations changes the qualitative aspects of the dynamics, see Fig. \ref{fig-vibNS}, showing the spatially resolved time-evolution of the molecular charge and spin from simulations with vibrating chiral molecules. First, it can be noticed that the vibrationally assisted spatial evolution is qualitatively similar to the evolution in the purely static molecule. However, while the charge in the static molecule returns to a nearly homogeneous  distribution in the stationary regime, Fig. \ref{fig-NS} (a), (e), in the vibrating molecule the charge distribution remains inhomogeneous acquiring a non-vanishing charge polarization also when the system dynamics is slowed down, approaching the stationary limit, see Fig. \ref{fig-vibNS} (a), (e), and inset of panel (a). Second, it can be noticed that the spin-polarizations that develop shortly after contact with the metal, also remain upon approaching the stationary state, Fig. \ref{fig-vibNS} (b)--(d), (f)--(h). While the transverse projections are configured anti-ferromagnetically within the molecule, Fig. \ref{fig-vibNS} (b), (c), (f), (g), the longitudinal projection converges towards a spin-polarized state with opposite polarizations at the two ends of the molecule, Fig. \ref{fig-vibNS} (d), (h). The state that is eventually reached for the molecule in the regime with slow dynamics, corroborates the results obtained in the stationary regime \cite{NanoLett.21.3026}, albeit the two approximations used are not exactly equivalent. The consistency between the results, nonetheless, suggests that the dissipative component provided through the molecular vibrations should be realistic and sound.


While the property that the two enantiomers acquire opposite spin-polarization when in contact with the metal is expected, in the next step it will be shown that this feature also opens up for enantiomer separation. In experiments, this can be done by contacting the enantiomers on a ferromagnetic surface and measure, e.g. the adsorption rate \cite{Science.360.1331} or the force required to pull the molecule off the surface \cite{AdvMater.31.1904206}. Here, the enantiomer selectivity is studied through the magnetic properties of the composite system comprising both the molecule and the metal.

\begin{figure}[t]
\begin{center}
\includegraphics[width=\columnwidth]{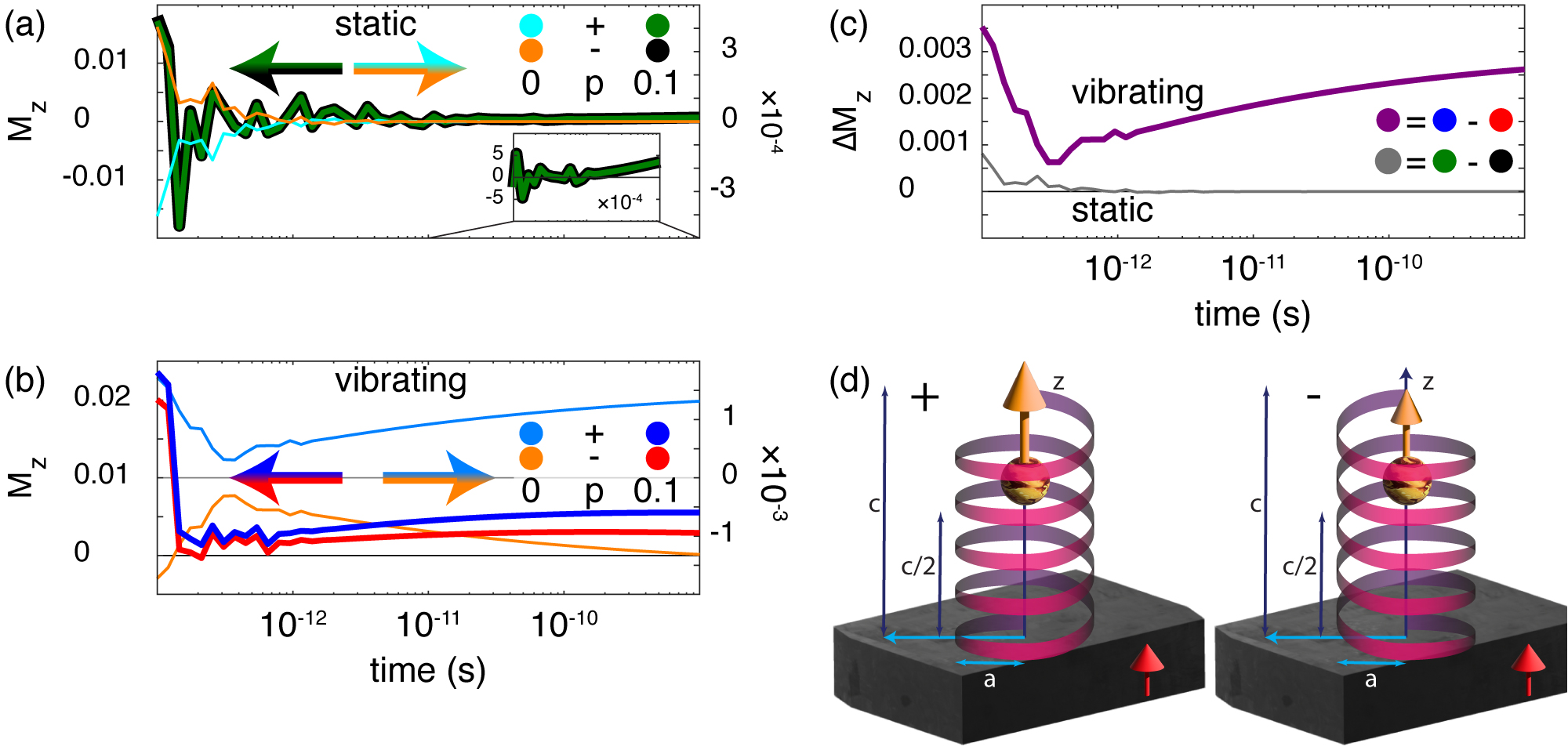}
\end{center}
\caption{(a) -- (c) Time evolution of mean spin-polarization $M^z_\pm$ for (a) static molecule, (b) vibrating molecule, and (c) differences $\Delta M_z$ between the enantiomers in the static and vibrating configurations. Bold (faint) lines in panels (a) and (b) represent the results in presence of ferromagnetic (normal) metal with $p=0.1$ ($p=0$), for the two enantiomers $\pm$. In panel (c), the bold (faint) line represent the vibrating (static) configurations.
(d) Schematic illustration of the chiral molecule when attached to the metal, defining the origin at the molecule-metal interface. The drawings also illustrate the induced mean spin-polarization immediately after contacting the molecule to the ferromagnetic metal. The magnetic moment of the metal is indicated with the arrow. The mean spin-polarization in the molecule with positive helicity is larger than that in the molecule with negative.
Other parameters are as in Fig. \ref{fig-vibNS}.}
\label{fig-polarizations}
\end{figure}

Using the spin-polarization as a tool for enantiomer separation was also considered in recent theoretical studies \cite{NanoLett.20.7077,JPhysChemLett.12.10262}. However, quantifying the enantiomer specific properties in terms of, e.g., the transmission, which is essentially not a measurable quantity, introduces ambiguities in how to comprehend the results. The reason is that measurable quantities, such as the magnetic moment and charge current, are integrated over many degrees of freedom, in particular the energy dependencies of the involved mechanisms. Therefore, it is questionable whether the detailed spectral properties are meaningful in a context where such cannot be resolved.

Hence, here the enantiomer specific properties are investigated in terms of the mean spin-polarization $M^z=2\sum_m(\av{z}-z_m)\av{s_m^z}/c$, where $\av{z}=\sum_mz_m/\mathbb{M}$ and the origin located at the molecule-metal interface, which defines a measure using which it is possible to determine whether there is any difference in the magnetic moments between the two enantiomers. As such, a positive (negative) spin-polarization can be understood as spin $\up$ ($\down$) being accumulated towards the interface with the metal accompanied with a spin $\up$ ($\down$) depletion towards the free end of the molecule. The advantage with this measure is that it directly connects to integrated measurable quantities such as the total magnetic moment since it provides a relation between the local magnetic moments $\av{s_m^z}$ and the overall structure.

The mean spin-polarizations of static and vibrating molecules when in contact with a metal, are shown in Fig. \ref{fig-polarizations}. Bold (faint) lines in panels (a) and (b) correspond to simulations of the molecule in contact with a ferromagnetic (normal) metal with $p=0.1$ ($p=0$), while the plots in panel (c) show the difference $\Delta M^z=M^z_+-M^z_-$, where $M^z_\pm$ denotes the spin polarization for the $\pm$ enantiomers, for the vibrating (bold) and static (faint) configurations. Although the quantitative details vary between the static and vibrating configurations, the overall conclusion that can be drawn is that the enantiomers acquire different mean spin-polarizations, particularly in the transient regime. While the static configuration does not allow for a distinction between the enantiomers when approaching the stationary limit, the vibrating configurations acquire a finite stationary mean spin-polarization. Of great importance for the chiral induced spin selectivity effect, however, is that the two enantiomers acquire different amplitudes of their spin-polarizations also in the stationary limit.

The results of the simulations can be interpreted as the schematic illustrations in Fig. \ref{fig-polarizations} (d), where the ball-arrow suggest the mean spin-polarization of the $+$ (left) and $-$ (right) enantiomers when in contact with the ferromagnetic substrate. Hence, the chirality of the $+$ ($-$) enantiomer cooperates (counteracts) the ferromagnetism in the substrate, leading to a enhanced (reduced) mean spin-polarization as the system approaches the stationary regime.


Here it should be stressed that chiral induced spin selectivity does not originate from the emergence of a spin-polarization in the molecule in itself, but it is the unequal amplitudes of the spin-polarizations of the enantiomers that form the basis for the phenomenon. This statement can be understood from the following discussion. The chiral induced spin selectivity effect is founded on the difference of the charge currents through a chiral molecule under different conditions. It can, for instance, be electrons photoexcited by circularly polarized light with opposite helicity such that the two photocurrents are different \cite{Science.331.894}. While there is no question that the environment has a strong effect on the magnetic properties of the molecule, however, unless there is an immanent property of the molecule that responds differently when the light helicity is changed, there cannot be any change in the spin-polarization of the photoelectric current. In this sense, the emergence of the spin-polariztion is intimately related to the excited states of the molecule. In the transient regime, these excitations are made available by the dynamical changes of the electronic structure, however, when approaching the stationary regime, those excitations are no longer accessible in the static configuration considered here. That is, in the static configuration there is no mechanism that allows for transferring electron density between states and, in particular, there is no mechanism that sustains an angular momentum transfer within the molecule. Nuclear, or, molecular vibrations, by contrast, facilitate transfer of both electron density and angular momentum between states, or, channels in the molecule. Moreover, at room temperature there is a wide energy window available, about 26 meV, for electrons to transfer between states through thermal excitations, which is a reason for the effectiveness of the nuclear vibrations in this context.

The assumption of instantaneous attachment of the molecule to the metal is crude and may cause unrealistically large charge transfer within the molecule in the simulations. However, since the approximation can be regarded as a limiting process of any non-adiabatic attachment to the metal, it does provide the conceptual mechanism in this context.


In summary, it has been shown that spin-polarization is dynamically generated in chiral molecules upon interfacing it with a metal. The dynamics indicate that chiral induced spin selectivity is an excited states phenomenon which, in the transient regime is sustained by the strong fluctuations imposed by the temporally changing conditions. When approaching the stationary regime, the system drives towards a ground state which may be spin-polarized, however, the existence of a non-vanishing spin-polarization depends on whether electronic density and angular momentum can be transferred between the states in the ground state. As has been shown here, nuclear vibrations do allow for such transfer and that these result provide fundamental clues for further development of a comprehensive theory for enantiomer selectivity.

\section{Acknowledgments}
The author thanks R. Naaman for constructive and fruitful discussions. Support from Vetenskapsr\aa det and Stiftelsen Olle Engkvist Byggm\"astare is acknowledged.

\end{document}